\documentclass[usenatbib]{mn2e}
\usepackage{epsfig} \usepackage{natbib} \usepackage{amssymb}
\usepackage{amsmath}

  \newcommand{\apj}{Astroph. J.}
\newcommand{\apjs}{Astroph. J. Suppl. Ser.}
  
\newcommand{\mnras}{Month. Notic. R. Astron. Soc.}

\newcommand{\aj}{Astron. J.}  \newcommand{\apjl}{Astroph. J. L.}

\newcommand{\HI}{H{\scriptsize{\textrm{I}}}}
\newcommand{\HII}{H{\scriptsize{\textrm{II}}}}

\newcommand{\HeII}{He{\scriptsize{\textrm{II}}}}
\newcommand{\HeIII}{He{\scriptsize{\textrm{III}}}}
\newcommand{\Lylim}{Lyman limit}
\newcommand{\fstar}{$f^*_{{\mathrm{esc}}}$}
 \newcommand{\zsix}{$z \sim
6$} \newcommand{\zol}{$z_{\mathrm{ol}}$}

\newcommand{\nul}{$\nu_{\scriptscriptstyle{\mathrm{L}}}$}
\newcommand{\nuz}{$\nu_{\scriptscriptstyle\mathrm{z}}$}
\newcommand{\mathnul}{\nu_{\scriptscriptstyle\mathrm{L}}}

\newcommand{\deltacrit}{$\Delta_{\mathrm{{c}}}$}
\newcommand{\lya}{Ly$\alpha$}

\def\simgt{\mathrel{\spose{\lower 3pt\hbox{$\sim$}} \raise
2.0pt\hbox{$>$}}} \def\simlt{\mathrel{\spose{\lower
3pt\hbox{$\sim$}}\raise 2.0pt\hbox{$<$}}}

\title[Reionization Bias in Quasar Near-zones]{Reionization Bias in High Redshift Quasar Near-Zones}

\author[Wyithe, Bolton \& Haehnelt]{J. Stuart B. Wyithe$^1$, James
S. Bolton$^2$ \& Martin G. Haehnelt$^3$\\ $^1$School of Physics,
University of Melbourne, Parkville, Victoria, Australia\\ $^2$Max
Planck Institut fur Astrophysik, Karl-Schwarzschild Str. 1, 85748
Garching, Germany\\ $^3$Institute of Astronomy, University of
Cambridge, Madingley Road, Cambridge, CB3 0HA\\ Email:
swyithe@physics.unimelb.edu.au}

\date{Draft Version}
\pagerange{\pageref{firstpage}--\pageref{lastpage}} \pubyear{2006}

\def\LaTeX{L\kern-.36em\raise.3ex\hbox{a}\kern-.15em
    T\kern-.1667em\lower.7ex\hbox{E}\kern-.125emX}

\begin{document}

\label{firstpage}

\maketitle

\begin{abstract} 

Absorption spectra of high redshift quasars exhibit an increasingly
thick Ly$\alpha$ forest, suggesting the fraction of neutral hydrogen
in the intergalactic medium (IGM) is increasing towards $z\sim6$.
However, the interpretation of these spectra is complicated by the
fact that the Ly$\alpha$ optical depth is already large for neutral
hydrogen fractions in excess of $10^{-4}$, and also because quasars
are expected to reside in dense regions of the IGM.  We present a
model for the evolution of the ionization state of the IGM which is
applicable to the dense, biased regions around high-redshift quasars
as well as more typical regions in the IGM.   We employ a cold dark
matter (CDM) based model in which the ionizing photons for
reionization are produced by star formation in dark matter halos
spanning a wide range of masses, combined with numerical radiative
transfer simulations which model the resulting opacity distribution in
quasar absorption spectra.  With an appropriate choice for the
parameter which controls the star formation efficiency, our model is
able to simultaneously reproduce the observed \lya~ forest opacity at
$4<z<6$, the ionizing photon mean-free-path at $z\sim4$ and the rapid
evolution of highly ionized near-zone sizes around high-redshift
quasars at $5.8<z<6.4$.  In our model, reionization extends over a
wide redshift range, starting  at $z \ga10$  and completing as HII
regions overlap at  $z\sim 6-7$. We find that within 5 physical Mpc of
a high redshift quasar, the evolution of the ionization state of the
IGM precedes that in more typical regions by around 0.3 redshift
units. More importantly, when combined with the rapid increase in the
ionizing photon mean-free-path expected shortly after overlap, this
offset results in an ionizing background near the quasar which exceeds
the value in the rest of the IGM by a factor of $\sim2-3$. We further
find that in the post-overlap phase of reionization the size of the
observed quasar near-zones is not directly sensitive to the neutral
hydrogen fraction of the IGM. Instead, these sizes probe the level of
the background ionization rate and the temperature of the surrounding
IGM.  The observed rapid evolution of the quasar near-zone sizes  at
$5.8<z<6.4$ can thus be explained by the rapid evolution of the
ionizing background, which in our model is caused by the completion of
overlap at the end of reionization by $6\la z\la7$.
\end{abstract}
    

\section{{INTRODUCTION}}
\label{introduction}

The reionization of cosmic hydrogen, which is commonly believed to
have been due to ultraviolet photons produced by the first stars and
quasars (Barkana \& Loeb~2001), was an important milestone in the
history of the Universe.  The recent discovery of very distant quasars
has enabled detailed studies of the ionization state of the high
redshift IGM to be made at a time when the universe was less than a
billion years old (Fan et al.~2006; White et al.~2003).  Several
studies have used the evolution of the ionizing background inferred
from these spectra to argue that the reionization of cosmic hydrogen
was completed just beyond {\zsix} (White et al.~2003; Fan et al.~2006;
Gnedin \& Fan~2006). However other authors have claimed that the
evidence for this rapid change  becomes significantly weaker for a
different choice density distribution in the IGM (Becker et
al~2007). One reason for the ambiguity in interpreting these
absorption spectra is that \lya~ absorption can only be used to probe
neutral fractions that are larger than $10^{-4}$, owing to the large
cross-section of the \lya~ resonance.

Several of the most distant quasars exhibit complete Gunn-Peterson
troughs, the red edges of which do not extend as far as the redshifted
Ly$\alpha$ wavelength.  A possible interpretation of the observation
of a Gunn-Peterson trough which does not extend all the way to the
Ly$\alpha$ wavelength is the presence of an \HII~ region in a partially
neutral IGM (e.g. Cen \& Haiman~2000, Madau \& Rees~2000). This
interpretation has been used in different ways to argue that there is
a significant fraction of neutral hydrogen in the IGM beyond $z\sim6$
(Wyithe \& Loeb~2004; Mesinger \& Haiman~2004; Wyithe, Loeb \&
Carilli~2005), with a lower limit on the hydrogen neutral fraction
that is as much as two orders of magnitude larger than constraints
available from direct absorption studies. Bolton \& Haehnelt (2007a),
Maselli et al. (2007) and Lidz et al. (2007), however, have argued
that  this interpretation is uncertain and that these observed regions
of transmission could either be due to an \HII~ region or due to a
classical proximity zone.

Given this uncertainty, and in order to connect the evolution of the
IGM ionization state at redshifts both above and below those where the
Gunn-Peterson troughs have been observed, Fan et al.~(2006) proposed
defining the sizes of these regions of transmission as the distance
where the smoothed spectrum first drops below the level where 10 per
cent of the flux is transmitted.  Bolton \& Haehnelt (2007a) refer to
this region as the highly ionized near-zone, and Fan et al.~(2006)
argued that its size should reflect the ionization state of the IGM in
a simple way. Using this definition, Fan et al.~(2006) found a rapid
trend of near-zone size with redshift, from which they inferred an
order of magnitude increase in the neutral fraction of the IGM over
the observed range of $5.8\la z\la6.4$.  However Bolton \&
Haehnelt~(2007a) demonstrated that the near-zone size  becomes
independent of the neutral hydrogen fraction when the IGM is highly
ionized.

In this paper we shall address two points regarding the interpretation
of the near-zones observed in  high redshift quasar absorption
spectra. As pointed out recently by Lidz et al.~(2007) and Alvarez \&
Abel (2007), the quasars observed prior to $z\sim6$ are likely to have
formed in regions substantially preionized by nearby clustered
galaxies.  We therefore first discuss the extent to which regions of
the IGM observed close to high redshift quasars in absorption spectra
are representative of the Universe on average, using a density
dependent semi-analytic model for the reionization history.  This
model is carefully calibrated to be consistent with the current
observational data, and as such provides a consistent theoretical
framework for modeling quasar near-zones in highly biased regions of
the IGM.  We then use this model to predict the sizes of quasar
near-zones and their evolution with redshift, both semi-analytically
and using detailed simulations of radiative transfer.

We begin by describing our density dependent semi-analytic model for
the reionization history (\S~\ref{models}), including our procedures
for the calculation of the ionizing background. We then present results
for the bias of reionization near quasars in \S~\ref{results}, before
discussing the evolution of the observed high redshift quasar
near-zones using the semi-analytic model in \S~\ref{nearzones}.  We
then explore this model in more detail with full radiative transfer
simulations in \S~\ref{sec:sims}. We present our conclusions in
\S~\ref{conclusions}. Throughout the paper we adopt the set of
cosmological parameters determined by {\it WMAP} (Spergel et al. 2007)
for a flat $\Lambda$CDM universe.

\section{Semi-Analytic Model for Reionization}
\label{models}

Miralda-Escude et al.~(2000) presented a model which allows the
calculation of an effective recombination rate in an inhomogeneous
universe by assuming a maximum overdensity ({\deltacrit}) which may be
penetrated by ionizing photons within {\HII~} regions. The model
assumes that reionization progresses rapidly through islands of lower
density prior to the overlap of individual ionized regions. Following
overlap, the remaining regions of high density are gradually
ionized. It is therefore hypothesised that at any time, regions with
gas below some critical normalised density threshold $\Delta_{\rm
i}\equiv {\rho_{i}}/{\langle\rho\rangle}$ are ionized while regions of
higher density are not. In what follows, we draw primarily from their
prescription and refer the reader to the original paper for a detailed
discussion of its motivations and assumptions.  Wyithe \& Loeb~(2003)
employed this prescription within a semi-analytic model of
reionization. In the current work we limit our attention to
reionization due to population-II stars (Gnedin \& Fan~2006;
Srbinovsky \& Wyithe~2006), which govern the final stages of
reionization even in the presence of an earlier partial or full
reionization by population III stars (e.g. Wyithe \& Loeb~2003).

Within the model of Miralda-Escude et al.~(2000) we describe the
post-overlap evolution of the IGM by computing the evolution of
${dF_{\rm M}(\Delta_{\rm i})}/{dz}$, where
\begin{equation}
F_{\rm M}(\Delta_{\rm i})=\int_{0}^{\Delta_{\rm i}}d\Delta P_{\rm
V}(\Delta)\Delta
\end{equation}
is the fraction of mass in regions with overdensity below
$\Delta_{\rm i}$, and $P_{\rm V}(\Delta)$ is the volume weighted
probability distribution for $\Delta$. In a region of large scale over
density $\delta$ at $z_{\rm obs}$, the mass fraction $F_{\rm
M}(\Delta_{\rm i})$ (or equivalently $\Delta_{\rm i}$) therefore
evolves according to the equation
\begin{equation}
\label{postoverlap}
\frac{dF_{\rm M}(\Delta_{\rm i})}{dz} =
\frac{1}{n_0}\frac{dn_\gamma(\delta)}{dz}-\alpha_{\rm
B}\frac{R(\Delta_{\rm i})}{a^3}n_{\rm
e}\left(1+\delta\frac{D(z)}{D(z_{\rm obs})}\right)\frac{dt}{dz},
\end{equation}
where $D$ is the growth factor, $\alpha_{\rm B}$ is the case B
recombination coefficient, $n_{\rm e}$ is the comoving electron
density, and $R(\Delta_{\rm i})$ is the effective clumping factor of
the IGM (see below).  This equation was described in Wyithe \&
Loeb~(2003), but it is generalised in this work to apply to regions of
large scale overdensity $\delta$ that differ from the average IGM.

Integration of equation~(\ref{postoverlap}) requires knowledge of
$P_{\rm V}(\Delta)$.  Miralda-Escude et al.~(2000) found that a good
fit to the volume weighted probability distribution for the density as
seen in cosmological hydrodynamic simulations has the functional form
\begin{equation}
P_{\rm
V}(\Delta)d\Delta=A\exp{\left[-\frac{(\Delta^{-2/3}-C_0)^2}{2(2\delta_0/3)^2}
\right]}\Delta^{-\beta}d\Delta,
\end{equation}
with $\delta_0=7.61/(1+z)$ and $\beta=2.23$, 2.35 and 2.48, and
$C_0=0.558$, 0.599 and 0.611 at $z=2$, 3 and 4. At $z=6$ they assume
$\beta=2.5$ and solve for $A$ and $C_0$ by requiring the mass and
volume to be normalised to unity. We repeat this procedure to find
$P_{\rm V}(\Delta)$ at higher redshifts. The proportionality of
$\delta_0$ to the scale factor is expected for the growth of structure
in an $\Omega_{\rm m}=1$ universe or at high redshift otherwise, and
its amplitude should depend on the amplitude of the
power-spectrum. The simulations on which the distribution in
Miralda-Escude et al.~(2000) was based assumed $\Omega_{m}=0.4$,
$\Omega_\Lambda=0.6$ and $\sigma_8=0.79$, close to the values used in
this paper. The above probability distribution remains a reasonable
description at high redshift when confronted with a more modern
cosmology and updated simulations, although the addition of an
analytical approximation for the high density tail of the
distribution remains necessary as a best guess at correcting for
numerical resolution (Bolton \& Haehnelt~2007b).

Equation~(\ref{postoverlap}) provides a good description of the
evolution of the ionization fraction following the overlap of
individual ionized bubbles because the ionization fronts are exposed
to the mean ionizing radiation field. However prior to overlap, the
prescription is inadequate due to the large fluctuations in the
intensity of the ionizing radiation. A more accurate model to describe
the evolution of the ionized volume prior to overlap was suggested by
Miralda-Escude et al.~(2000). In our notation the appropriate equation
is
\begin{eqnarray}
\nonumber \frac{d[Q_{\rm i}F_{\rm M}(\Delta_{\rm c})]}{dz} &=&
\frac{1}{n^0}\frac{dn_{\gamma}(\delta)}{dz}\\ &&\hspace{-25mm}-
\alpha_{\rm B}(1+z)^3R(\Delta_{\rm crit})n_{\rm
e}\left(1+\delta\frac{D(z)}{D(z_{\rm obs})}\right)Q_{\rm
i}\frac{dt}{dz}.
\end{eqnarray} 
or
\begin{eqnarray}
\label{preoverlap}
\nonumber \frac{dQ_{\rm i}}{dz} &=& \frac{1}{n^0 F_{\rm M}(\Delta_{\rm
crit})}\frac{dn_{\gamma}(\delta)}{dz}\\ \nonumber &-&\left[\alpha_{\rm
B}(1+z)^3R(\Delta_{\rm crit})n_{\rm
e}\left(1+\delta\frac{D(z)}{D(z_{\rm
obs})}\right)\frac{dt}{dz}\right.\\ &&\hspace{15mm}+
\left.\frac{dF_{\rm M}(\Delta_{\rm crit})}{dz}\right]\frac{Q_{\rm
i}}{F_{\rm M}(\Delta_{\rm c})}.
\end{eqnarray} 
In this expression, $Q_{\rm i}$ is redefined to be the volume filling
factor within which all matter at densities below $\Delta_{\rm c}$ has
been ionized. The effective clumping of the IGM can now be written
\begin{equation}
R(\Delta_{\rm i})\equiv Q_{\rm i}\frac{\langle \rho^2\rangle}{\langle
\rho\rangle^2} =\int_{0}^{\Delta_{\rm i}}d\Delta P_{\rm
V}(\Delta)\Delta^2,
\end{equation}
where $F(\Delta_{\rm i})$ is the fraction of the volume with
$\Delta<\Delta_{\rm i}$.

Within this formalism, the epoch of overlap is precisely defined as
the time when $Q_{\rm i}$ reaches unity. However, we have only a
single equation to describe the evolution of two independent
quantities $Q_{\rm i}$ and $F_{\rm M}$. The relative growth of these
depends on the luminosity function and spatial distribution of the
sources. The appropriate value of $\Delta_{\rm c}$ is set by the mean
separation of the ionizing photon sources. More numerous sources can
attain overlap for smaller values of $\Delta_{\rm c}$. We assume
$\Delta_{\rm c}$ to be constant with redshift before overlap with
values of $\Delta_{\rm c}=20$ and $\Delta_{\rm c}=5$ (which  lie in
the range for galaxies) in this paper, and show examples for both.

Our approach is to compute a reionization history given a particular
value of $\Delta_{\rm c}$, combined with assumed values for the
efficiency of star-formation and the fraction of ionizing photons that
escape from galaxies. With this history in place we then compute the
evolution of the background radiation field due to these same sources.
Post overlap, ionizing photons will experience attenuation due to
residual overdense pockets of {\HI} gas.  We use the description of
Miralda-Escude et al.~(2000) to estimate the ionizing photon
mean-free-path, and subsequently derive the attenuation of ionizing
photons. We then compute the flux at the {{\textit
\Lylim}}~({\nul}~$=3.29\times10^{15}\mathrm{Hz})$ in the IGM due to
sources immediate to each epoch, in addition to redshifted
contributions from earlier epochs. This calculation follows the model
described in Srbinovsky \& Wyithe~(2006) [see also Choudhury \& Ferrara~2005] but is summarised below.

We note that {\HI} {\Lylim} photons $(13.6\hspace{2pt}eV)$ are
incapable of ionizing helium (He) [$24.6\hspace{2pt}eV$ ({\HeII}),
$54.4\hspace{2pt}eV$ ({\HeIII})]. We therefore neglect He when
computing the intensity of the ionizing background in our
semi-analytic model.

\subsection{{Evaluating flux at the Lyman limit}}
\label{evaluating_flux}

In this section we review the process for calculating the ionizing
background flux at a particular redshift. We assume the post overlap
ionizing background to be generated by population-II stars alone
following the results of Srbinovsky \& Wyithe~(2006). We first define
$z_0$ to be the redshift at which the flux is to be evaluated. The
flux at the Lyman limit is normalised in \emph{physical} units of
$J_{21}$ $(10^{-21} \mathrm{ergs/sec/Hz/cm^2/sr})$, and at redshift
$z_0$ it is related to the energy density by

\begin{equation}
\label{dEdV_to_flux}
J_{21}(z_0) =
\frac{c}{4\pi}\hspace{2pt}\frac{d^2E_{\mathnul}^{{\mathrm{tot}}}(z_0)}{dV{d\nu}}\frac{1}{10^{-21}}
(1+z_0)^3,
\end{equation}

\noindent
where $c$ is the speed of light and $\frac{ d^2
E_{{\mbox{\scriptsize\nul}}}^{{\mathrm{tot}}}(z_0) } {dV{d\nu}}$ is
the energy per unit frequency interval per unit co-moving volume at
frequency {\nul} and redshift $z_0$.  Contributions to the radiation
field at {\nul} (and $z_0$) from sources at redshift $z$ were emitted
at frequency {\nuz}~ $= \frac{1+z}{1+z_0}${\nul}.  Note that {\nuz}
remains below the He {\Lylim} $(24.6\hspace{2pt} eV)$ whilst
$\frac{1+z}{1+z_0} \la 2$.  The semi-analytic model of the
reionization history predicts an average redshift of overlap, which we
refer to as {\zol}.

The ionizing background flux contains contributions from sources at
$z_0$ in addition to redshifted flux from sources at higher redshift.
To compute the ionizing background flux we consider contributions from
sources with redshifts $z>z_0$. The total co-moving energy density at
redshift $z_0$ is
{
\begin{equation}
\label{fluxeq}
\frac{d^2E_{{\mbox{\scriptsize\nul}}}^{\mathrm{tot}}(z_0)}{dVd\nu}=\int_{\infty}^{z_0}dz
\frac{d^3E_{{\mbox{\scriptsize\nuz}}}(z)}{dVd{\nu}dt} Q_{\rm
i}\left(\frac{1+z_0}{1+z}\right)^3  e^{-\tau(z,z_0)} \frac{dt}{dz},
\end{equation}
}

\noindent
The term $\frac{d^3E_{{\mbox{\scriptsize\nuz}}}(z)}{dVd{\nu}dt}$
represents the frequency dependent (co-moving) energy density per unit
time generated by ionizing sources, and $\tau$ is the optical depth
for ionizing photons between $z$ and $z_0$. Rather than
assume the complete attenuation of ionizing photons emitted prior to
the redshift of overlap, we instead assume attenuation due to the
mean-free-path evaluated in ionized regions prior to overlap and
weight the contribution to the energy density by $Q_{\rm i}$. This
approach should approximate the contribution of sources to the
ionizing background that emit at redshifts late in the overlap epoch
when the typical bubble size is evolving rapidly, and is much larger
than the ionizing photon mean-free-path (Furlanetto, Zaldarriaga \&
Hernquist~2004).

Finally, the photoionization rate may be related to the observed
ionizing background flux using
\begin{equation}
\label{Gamma}
\Gamma= 4\pi \int_{\mbox\nul}^{\infty}
\frac{J_{\nu}}{h_{\mathrm{p}}\nu} \sigma_{\nu} d\nu ,
\end{equation}
where $\sigma_{\nu}$ is the {\HI} photo-ionization
\textit{cross-section}  and $h_{\mathrm{p}}$ is Planck's constant.
Using equation~(\ref{Gamma}) and the spectra describing our sources,
we find\footnote{This conversion was reported incorrectly in
Srbinovsky \& Wyithe~(2006) as $\Gamma_{12}=1.69J_{21}$.}
$\Gamma_{12}=2.82J_{21}$ (where $\Gamma_{12}$ is the ionizing rate in
units of $10^{-12}\mathrm{s}^{-1}$) for a background in which stars
are the dominant source. This last step does not account for the
variation of the spectral shape (due to redshifting) from that emitted
at a constant time.

\subsection{{\Lylim} photons from Stars}
\label{stellar_flux}

We next describe the contribution to the ionizing background flux made
by population-II stars. To begin, we describe the spectral energy
distribution (SED) of population-II star forming galaxies, using the
model presented in Leitherer et al.~(1999). The SED has the form {
$\frac{d^3E_\nu}{d\nu dtd\dot{M}} $}, where {$\dot{M}$} is expressed
in units of (baryonic) solar masses per year. In the instance that
ionizing photons are produced primarily in star bursts, with lifetimes
much shorter than the Hubble time, we may express the star formation
rate per unit time as
\begin{equation}
\frac{d\dot{M}}{dV}(z)=f^*\frac{dF(\delta,z)}{dt_{\mathrm{year}}}\hspace{2pt}\rho_{\mathrm{b}},
\end{equation}
where $\rho_{\mathrm{b}}$ is the co-moving baryonic mass density, and
$F(\delta,z)$ is the density dependent collapsed fraction of mass in
halos above a critical mass at $z$. The factor  $f^*$ (the star formation
efficiency) describes the fraction of collapsed matter that
participates in star formation. This fraction is largely unknown.

In a region of co-moving radius $R$ and mean overdensity
$\delta(z)=\delta D(z)/D(z_{\rm obs})$ [specified at redshift $z$
instead of the usual $z=0$], the relevant collapsed fraction is
obtained from the extended Press-Schechter~(1974) model (Bond et
al.~1991) as
\begin{equation}
F(\delta,R,z) = \mbox{erfc}{\left(\frac{\delta_{\rm
c}-\delta(z)}{\sqrt{2\left(\left[\sigma_{\rm
gal}\right]^2-\left[\sigma(R)\right]^2\right)}}\right)},
\end{equation}
where $\mbox{erfc}(x)$ is the error function, $\sigma(R)$ is the
variance of the density field smoothed on a scale $R$, and
$\sigma_{\rm gal}$ is the variance of the density field smoothed on a
scale $R_{\rm gal}$, corresponding to a mass scale of $M_{\rm min}$ or
$M_{\rm ion}$ (both evaluated at redshift $z$ rather than at $z=0$).
In this expression, the critical linear overdensity for the collapse
of a spherical top-hat density perturbation is $\delta_c\approx 1.69$.

In a cold neutral IGM beyond the redshift of reionization, the
collapsed fraction should be computed for halos of sufficient mass to
initiate star formation. The critical virial temperature is set by
the temperature ($T_{{\mathrm{N}}}\sim 10^4$ K) above which efficient
atomic hydrogen cooling promotes star formation. Following the
reionization of a region, the Jeans mass in the heated IGM limits
accretion to halos above $T_{{\mathrm{I}}}\sim10^5$ K
(Efstathiou~1992; Thoul \& Weinberg~1996; Dijkstra et al.~2004).  We
may therefore write the time derivative of the collapsed fraction
\begin{align}
\label{collapsed_fraction}
\frac{dF}{ dt_{  {\mathrm{year}}  }}(z)= &  \left[
Q_{\mathrm{m}}(z)\frac{dF(z,T_{{\mathrm{I}}})}{dz} +
[1-Q_{\mathrm{m}}(z)] \frac{dF(z,T_{\mathrm{N}})}{dz}\right] \notag\\
& \times \frac{dz}{dt_{{\mathrm{year}}}},
\end{align}
where $Q_{\mathrm{m}}$ is the ionized mass fraction in the universe.

To describe the ionizing flux from stars we require one further
parameter. Only a fraction of ionizing photons produced by stars may
enter the IGM. Therefore an additional factor of $f_{\mathrm{esc}}$
(the escape fraction) must be included when computing the emissivity
due to stars.  Ciardi \& Ferrara~(2005) include a review of existing
constraints on $f_{\mathrm{esc}}$ which suggests that its value is
$\la 15$ per cent. The star formation efficiency and escape fraction
may be combined into a single free parameter ({\fstar}) to describe
the contribution of stars in our model.

Finally the energy density due to population-II stars at $z_0$ may be
computed using equation~(\ref{fluxeq}) with
$\frac{d^3E_{\nu_z}(z)}{dVd{\nu}dt}$  given by {
\label{stellar_energy_density}
\begin{equation}
\frac{d^3E_{\mbox\nuz}(z)}{dVd{\nu}dt}\hspace{2pt}  =
\frac{d^3E_{\mbox\nuz}(z)}{d\nu dt{d\dot{M}}}
\frac{dF(z)}{dt_{\mathrm{year}}}\hspace{2pt}\rho_{\mathrm{b}}
{\mbox\fstar}.
\end{equation}
}
\noindent This energy density may then be converted to a flux.

\subsection{\normalsize{Attenuation of ionizing photons}}
\label{attenuation}

Having computed a luminosity density we may now compute a value for
the ionizing background using equations
(\ref{dEdV_to_flux}-\ref{Gamma}), following the calculation of the
optical depth ($\tau$).  We fix {\deltacrit} prior to overlap, and
then follow the evolution of $\Delta_{\rm i}$ following overlap. In
each case we use the approach of Miralda-Escude et al.~(2000) to
estimate the mean-free-path ($\lambda$) as a function of redshift
$(z_i)$,

\begin{equation}
\label{Miralda-EscudeMFP}
{\lambda}_i ={\lambda}_0(1-F_{\mathrm{v}})^{-{2}/{3}}.
\end{equation}
Here
\begin{equation}
\label{volume_fraction}
F_{\mathrm{v}} = \int_0^{\Delta_{\rm max}(z)}d{\Delta} P({\Delta})
\end{equation}
is the volume filling factor of ionized regions, computed using the
the volume weighted probability distribution [$P(\Delta)$] of the
overdensity (Miralda-Escude et al.~2000), and $\Delta_{\rm max}$ is
equal to $\Delta_{\rm c}$ and $\Delta_{\rm i}$ prior to and following
overlap respectively. We note that prior to overlap the mean-free-path
is ill-defined, and we calculate the mean-free-path for photons in
{\em ionized regions} prior to overlap. The product ${\lambda}_0 H=60
\hspace{1pt}\mathrm{kms}^{-1}$ was obtained from comparison to the
scales of Ly$\alpha$ forest structures in simulations, at $z=3$
(Miralda-Escude et al.~2000). Finally the mean-free-path as a function
of redshift may be used to compute the attenuation of ionizing photons
between redshifts $z$ and $z_0$.  The resulting optical depth is
\begin{equation}
\label{optical_depth}
\tau(z,z_0) = \int^{z_0}_{z}dz \frac{cdt}{dz}\frac{1}{\lambda_i}.
\end{equation}

\subsection{The overdensity near massive halos}

Strong clustering of massive sources implies that these sources should
trace the higher density regions of IGM. In this section we compute
the distribution of overdensities on a scale $R$ that are centered on
halos of mass $M$. Our aim is to calculate the distribution of over
densities surrounding high redshift quasars, which are expected to be
larger compared to an average region of the IGM (e.g. Guimaraes et
al. 2007, Faucher-Gigurere et al. 2007).

The likelihood of observing a galaxy at a random location is
proportional to the number density of galaxies. At small values of the
large scale overdensity $\delta$, this density is proportional to
$[1+\delta b(M,z)]$, where $b$ is the galaxy bias (Mo \& White~1996;
Sheth, Mo \& Tormen~2002).  More generally, given a large scale over
density $\delta$ on a scale $R$, the likelihood of observing a galaxy
may be estimated from the Sheth-Tormen~(2002) mass function as
\begin{equation}
\label{LH}
\mathcal{L}_{\rm g}(\delta) = \frac{(1+\delta)\nu(1+\nu^{-2p})
e^{-a\nu^2/2}}{\bar{\nu}(1+\bar{\nu}^{-2p})e^{-a\bar{\nu}^2/2}},
\end{equation}
where $\nu=(\delta_{\rm c}-\delta)/[\sigma(M)]$ and $\bar{\nu} =
\delta_{\rm c}/[\sigma(M)]$. Here $\sigma(M)$ is the variance of the
density field smoothed with a top-hat window on a mass scale $M$ at
redshift $z$, and $a=0.707$ and $p=0.3$ are constants. Note that here
as elsewhere in this paper, we work with overdensities and variances
computed at the redshift of interest (i.e. not extrapolated to
$z=0$). Utilising Bayes theorem, we find the a-posteriori probability
distribution for the overdensity $\delta$ on the scale $R$ given the
locations defined by a galaxy population. We obtain
\begin{equation}
\left.\frac{dP}{d\delta}\right|_{\rm gal}\propto \mathcal{L}_{\rm
g}(\delta)\frac{dP_{\rm prior}}{d\delta},
\end{equation}

\noindent where $\frac{dP_{\rm prior}}{d\delta}$ is the prior
probability distribution for $\delta$, which is described by a
Gaussian of variance $\sigma(R)$. We may use this probability for the
overdensity within $R$ of a luminous high redshift quasar to calculate
the enhancement in the recombination and star-formation rates. Thus,
through integration of equations~(\ref{postoverlap}) and
(\ref{preoverlap}), we may compute the reionization history within
typical regions surrounding luminous quasars, and compare these with
the reionization history of the mean IGM ($\delta=0$).

\section{RESULTS }
\label{results}

\begin{figure*}
\includegraphics[width=14cm]{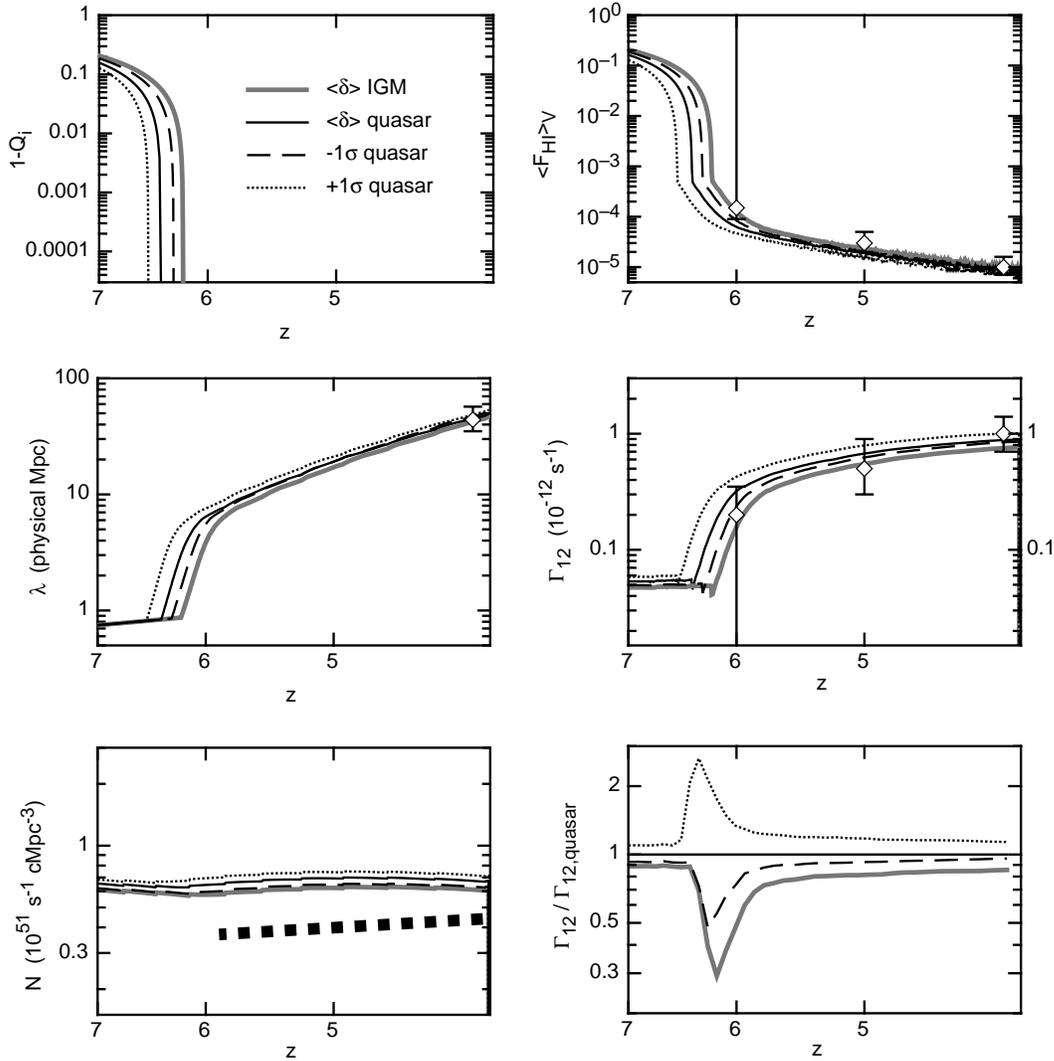}
\caption{\label{fig1}  The effect of overdensity on the redshift of
overlap, and the subsequent ionization state of the IGM. Four cases
are shown, corresponding to different overdensities evaluated in  5
physical Mpc spheres, including the average overdensity
centered on a quasar at $z\sim6$ (dark solid curves) and the plus and
minus 1-$\sigma$ fluctuations around that mean (dotted and dashed
curves). We also show the case of the mean IGM with $\delta=0$ (thick
grey lines).  {\em Top Left Panel:} The fraction of the IGM yet to
overlap according to the definition of overlap given in
\S~\ref{models}. {\em Top Right Panel:} The volume averaged filling
factor of neutral gas. {\em Central Left Panel:} The mean-free-path
for ionizing photons computed using the formalism in
\S~\ref{models}. {\em Central Right Panel:} The ionization rate.  {\em
Lower Left Panel:} The emissivity of ionizing sources. Also shown
(thick dotted line) is the observed evolution of ionizing sources as
estimated in Bolton \& Haehnelt~(2007b). {\em Lower Right Panel:} The
ratio of the ionization rate at different overdensities relative to
the average overdensity centered on a quasar host.  In this example the halo mass was $M=10^{13}M_\odot$ and $\Delta_{\rm c}=20$. The observational
points in the top and middle right panel are also from Bolton \&
Haehnelt~(2007b). }
\end{figure*}

The model described in the previous section has two free parameters
$\Delta_{\rm c}$ and {\fstar}. For any combination of these parameters
we are able to compute a reionization history, the evolution of the
mean-free-path, and the evolution of the ionizing background as a
function of the local overdensity. In this section we first summarise
the observations to which we compare our model before describing the
results of this comparison.

\subsection{Observational estimates of the ionization rate}

\begin{figure*}
\includegraphics[width=14cm]{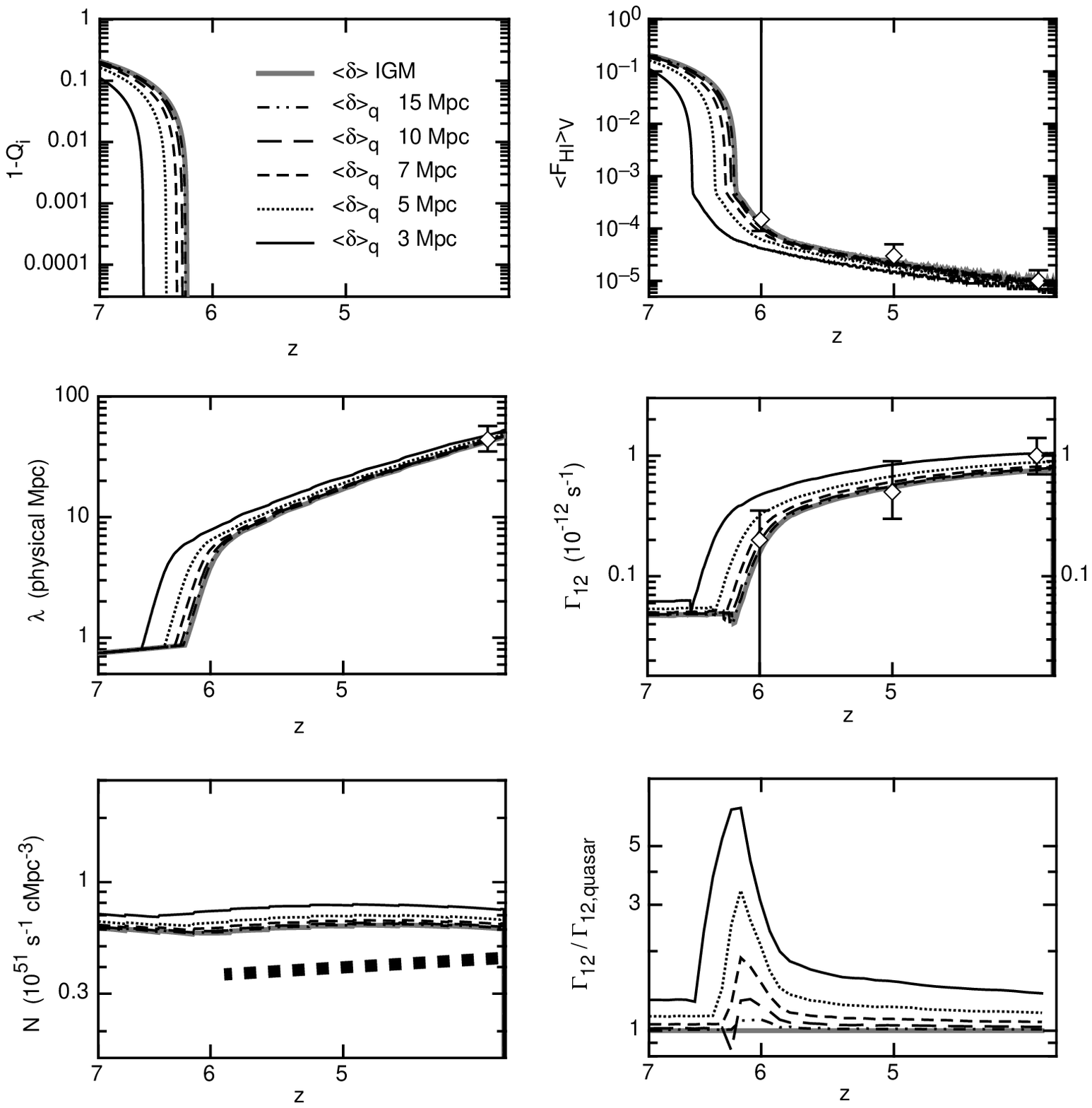}
\caption{\label{fig2} The effect of overdensity on the redshift of
overlap, and the subsequent ionization state of the IGM. Five cases
are shown, corresponding to overdensities evaluated within spheres
with radii between 5 and 15 physical Mpc, centered on a quasar
at $z\sim6$. In each case we evaluate the reionization history
assuming the mean overdensity surrounding the quasar. We also show
the case of the mean IGM with $\delta=0$ (thick grey lines).  {\em Top
Left Panel:} The fraction of the IGM yet to overlap according to the
definition of overlap given in \S~\ref{models}. {\em Top Right Panel:}
The volume averaged filling factor of neutral gas. {\em Central Left
Panel:} The mean-free-path for ionizing photons computed using the
formalism in \S~\ref{models}. {\em Central Right Panel:} The
ionization rate.  {\em Lower Left Panel:} The emissivity of ionizing
sources. Also shown (thick dotted line) is the observed evolution of
ionizing sources summarised in Bolton \& Haehnelt~(2007b). {\em Lower
Right Panel:} The ratio of the ionization rate at different over
densities relative to the average overdensity centered on a quasar
host.   In this example the halo mass was $M=10^{13}M_\odot$ and $\Delta_{\rm c}=20$. The observational points in the top and middle right panel
are also from Bolton \& Haehnelt~(2007b). }
\end{figure*}

Fan et al.~(2006) have analyzed the absorption spectra of 19 high
redshift quasars. Based on this data Bolton \& Haehnelt~(2007b)
present estimates of the ionization rate, $\Gamma_{12}$ and the
neutral hydrogen fraction over a range of redshifts near the end of
the reionization era using a suite of detailed numerical simulations.
In Figures~\ref{fig1} and \ref{fig2} we compare our model to these
estimates to demonstrate that it is consistent with the available
data.

\subsection{Comparison of reionization near a quasar and in the general IGM}
\label{comparison}

We now use the model developed earlier to address the level of bias in
reionization near to a luminous high redshift quasar relative to the
IGM as a whole.  We show results for a particular model in which we
assume that quasars are hosted by halos of mass $10^{13}M_\odot$ for
$\Delta_{\rm c}=20$. Although we compare our model for the ionizing
background to only three observational points at $z\ga4$ using two
free parameters, $\Delta_{\rm c}$ and {\fstar}, we find that these
parameters are not degenerate following overlap. Larger values of
$\Delta_{\rm c}$ increase the volume of hydrogen that must be
reionized prior to overlap, leading to a delay in the reionization
redshift. However, following overlap, the mean-free-path increases
rapidly, and hence $\Gamma_{12}$ at redshifts below $\sim5$ is quite
insensitive to $\Delta_{\rm c}$. On the other hand, varying {\fstar}
strongly affects the overlap redshift, since the critical number of
ionizing photons per baryon is achieved earlier. Following overlap, a
larger {\fstar} also increases the value of $\Gamma_{12}$, which is
proportional to the local emissivity in addition to the
mean-free-path.  We therefore adjust the parameter {\fstar} so that
the model evaluated at the mean density of the Universe gives a good
fit to the observed ionizing background (we find {\fstar}=0.0037). The
best fit to the ionization rate observed at $z\sim6$ is provided by a
value of $\Delta_{\rm c}\sim20$. Values of $\Delta_{\rm c}\ga5$ also
yield a satisfactory fit. The overlap redshifts corresponding to
($\Delta_{\rm c}$,{\fstar})= (20,0.0037) and ($\Delta_{\rm
c}$,{\fstar})= (5,0.0037) are $z_{\rm ol}=6.2$ and $z_{\rm ol}=7.2$
respectively.

Figure~\ref{fig1} shows the redshift evolution of $Q_{\rm i}$,
$\lambda_{\rm i}$, the volume averaged neutral fraction and the
ionization rate for the best fit model. Reionization histories for
four different values of overdensity were computed. These
overdensities were evaluated in 5 physical Mpc spheres, and correspond
respectively to the mean overdensity centered on a quasar at $z\sim6$
(dark solid lines), and to the plus and minus 1-$\sigma$ fluctuations
around that mean (dotted and dashed lines). We also show the case of
the mean IGM with $\delta=0$ (thick grey lines).  The examples
demonstrate the level of enhancement in the reionization process
within 5 physical Mpc of a high redshift quasar, and the expected
scatter around this enhancement.

In the top left panel of Figure~\ref{fig1} we show the fraction of the
IGM yet to overlap according to the definition given in
\S~\ref{models}. The more overdense regions experience overlap first
due to the enhanced production of ionizing photons in these
regions. In this model the mean IGM achieves overlap at
$z\sim6.1$. However within 5 physical Mpc of a quasar the overlap is
achieved $\sim0.25$ redshift units earlier on average, with a
1$-\sigma$ scatter of $\sim0.2$ redshift units.  These results are in
quantitative agreement with the analytic calculation of Lidz et
al.~(2007; see their Figure~1). However our model allows the
calculation of post-overlap properties such as the volume averaged
neutral fraction, ionizing photon mean-free-path and ionization rate
in addition to the pre-overlap mass-averaged ionization fraction.

In the central left panel of Figure~\ref{fig1} we show the
corresponding value of the mean-free-path for ionizing photons,
computed using the formalism in \S~\ref{models}. The model predicts a
rapid increase in the mean-free-path following overlap, which is also
indicated from the results of numerical simulations, (e.g. Gnedin
2000). However, the offset in the overlap redshift induced by large
scale overdensity leads to substantial differences in mean-free-path
within regions of different average overdensities for a brief period
following overlap. Once the rate of change of mean-free-path slows,
then its dependence on overdensity is reduced.

In the central right panel of Figure~\ref{fig1} we show the dependence
of the ionization rate on redshift. The rapid change in the
mean-free-path following overlap leads to a corresponding rapid change
in the ionization rate. Thus, for a brief period near overlap the
model predicts large variations in the ionizing background within
regions with different mean overdensities.  Note that the rapid
increase in $\Gamma_{12}$ is entirely due to the increasing
mean-free-path, and not to an increase in the source emissivity which
is shown in the lower left panel of Figure~\ref{fig1}. Indeed, the
radiative feedback that accompanies the end of the reionization era
results in a slight reduction of the emissivity during the time when
the ionizing background undergoes a large increase. For comparison the
fit to the observationally derived ionizing emissivity from Bolton \&
Haehnelt~(2007b) is shown by the thick dotted line. The model
emissivity is close to the observationally derived estimate (with the
offset due to the assumption of a somewhat harder spectrum here), and
it has a similar (small) redshift dependence.  In the upper right
panel we show the volume averaged hydrogen neutral fraction. Again,
near overlap the value of the neutral fraction inferred within the
vicinity of high redshift quasars is significantly biased relative to
the average IGM.  To better illustrate the level and duration of the
bias introduced by early overlap we also plot the ionization rate at
different overdensities relative to the average overdensity centered
on a quasar host halo (lower right hand panel of
Figure~\ref{fig1}). We see that at times near overlap, the ionizing
background within 5 physical Mpc of a quasar can be as much as a
factor of 3 in excess of the average ionizing background.  We have
also checked the enhancement of the ionizing background (at a fixed
distance) near a quasar host following overlap for different values of
the quasar host mass and $\Delta_{\rm c}$.  We find the enhancement is
reduced for smaller values of the quasar host mass, but is not
sensitive to the value of the critical over density $\Delta_{\rm c}$.

\subsection{The dependence of reionization on the distance from a quasar}

In the previous section we saw that the early reionization of
overdense regions will lead to an ionizing background within 5
physical Mpc of a high redshift quasar that is in excess of the
average IGM by a factor of $\sim3$. In this section, we explore the
variation of this enhancement with distance from the quasar. As
before, in Figure~\ref{fig2} we plot the effect of overdensity on the
redshift of overlap, and the subsequent ionization state of the
IGM. We again assume the case of $\Delta_{\rm c}=20$ and a quasar host
mass of $10^{13}M_\odot$. Five cases are shown, corresponding to
overdensities evaluated within spheres with radii between 5 and 15
physical Mpc, centered on a quasar at $z\sim6$. In each case we
evaluate the history assuming the mean overdensity surrounding the
quasar host of mass $M=10^{13}M_\odot$. We also show the case of the
mean IGM with $\delta=0$ (thick grey lines).

In the top left panel of Figure~\ref{fig2} we show the fraction of the
IGM that has yet to overlap. Reionization will occur earlier in
regions that are closer to the quasar, where the overdensity is
typically higher. In the central left panel we plot the mean-free-path
for ionizing photons. At a fixed time around overlap we expect that
the mean-free-path will be larger closer to the quasar. It is
important to note that throughout the epoch where the evolution of the
mean-free-path is rapid, the mean-free-path is smaller than the
distances from the quasar under consideration.  In the upper right and
central right hand panels we plot the volume averaged neutral fraction
of hydrogen and the ionization rate. As before we also plot the
model and observed emissivity of ionizing sources in the lower left
panel. Finally, in the lower right hand panel we show the ratio of the
average ionization rate within different radii relative to the
the value calculated at the average overdensity centered on a quasar
host.  As expected, we find that the level of enhancement decreases as
the distance from the quasar increases.

\begin{figure}
\includegraphics[width=8.25cm]{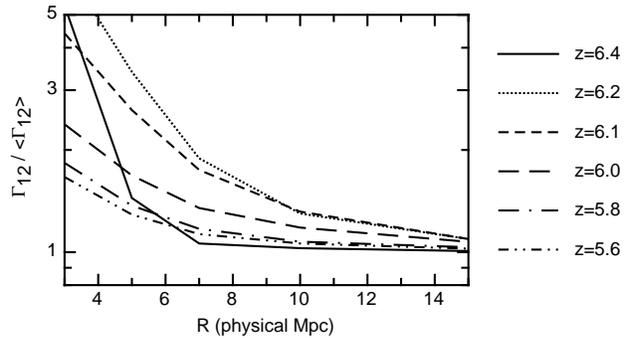}
\caption{\label{fig3} The ratio of ionization rates due to galaxies
calculated at the mean overdensity found within spheres of radius $R$
surrounding a high redshift quasar, relative to the ionization rate in
the mean IGM.  The curves are plotted over a range of redshifts
between $z=5.6$ and $z=6.4$. The overlap of the mean IGM in this model
(with $\Delta_{\rm c}=20$) is at $z=6.2$. The quasar host masses were
$10^{13}M_\odot$. }
\end{figure}

To better see the dependence of the ionization rate on radius, we plot
its enhancement within a spherical region of radius $R$ surrounding a
quasar relative to the mean IGM (Figure~\ref{fig3}).  The values are
shown for a range of redshift bins near overlap. The enhancement in
the ionizing background is greatest close to the quasar, where the
overdensity is largest. We find that the enhancement drops below 10
per cent beyond distances of 15 physical Mpc.  At all scales the
enhancement is largest at a time shortly after the overlap of the
average IGM. The enhancement more than halves in value by around half
a redshift unit after overlap, and has almost disappeared by 1
redshift unit following overlap.

Before proceeding we note that we have not included the effect of the
quasar in our calculation of the enhancement of the ionizing
background near the quasar. The additional ionizing flux from the
quasar would increase the local value of the ionizing photon
mean-free-path, and therefore increase the value of the ionization
rate due to the stellar emissivity. The enhancements described in
Figure~\ref{fig2} should therefore be considered as lower limits.

\section{The evolution of quasar near-zones}

\label{nearzones}

\begin{figure*}
\includegraphics[width=15cm]{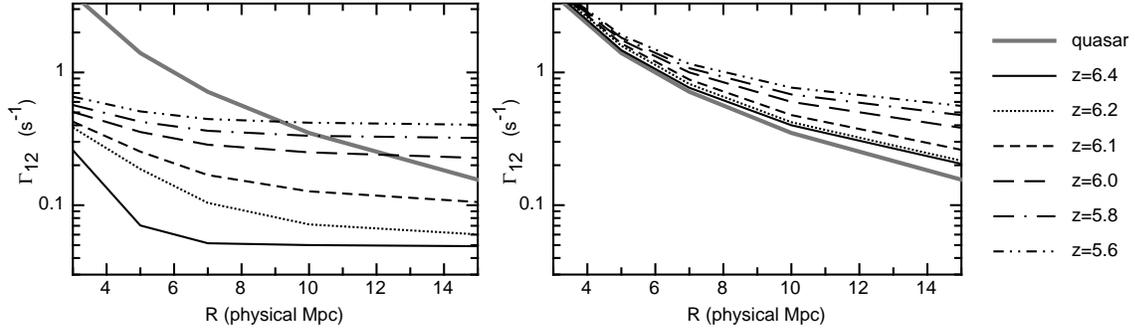}
\caption{\label{fig4} The ionization rate as a function of radius from
a quasar. {\em Left Panel:} The ionization rate due to
galaxies calculated at the mean overdensity found within spheres of radius
$R$ surrounding a high redshift quasar.
The curves are plotted at a range of redshifts between
$z=5.6$ and $z=6.4$. The ionization rate due to the quasar in a highly
ionized IGM is also shown (thick grey line). {\em Right Panel:} The
corresponding ionization rates due to galaxies plus the central
quasar. The overlap of the mean IGM in this model (with $\Delta_{\rm
c}=20$) is at $z=6.2$. The quasars host masses were $10^{13}M_\odot$,
and their ionizing luminosities were $\dot{N}=2\times10^{57}$s$^{-1}$.}
\end{figure*}

We now turn to the prediction of \lya~ near zone sizes.  Bolton \&
Haehnelt~(2007a), Maselli et al. (2007) and Lidz et al.~(2007) have
shown that the interpretation of near-zone sizes with respect to the
neutral hydrogen fraction in the IGM is difficult. Indeed, the data
may approximately correspond to the edge of an HII region surrounded
by a neutral IGM or instead correspond to a classical proximity zone
in an highly ionized IGM.  As discussed in the introduction, Fan et
al.~(2006) proposed defining the sizes of these regions of
transmission as the distance where the smoothed spectrum first drops
below the level where 10 per cent of the flux is transmitted.  Using
this definition, Fan et al.~(2006) found a rapid trend of near-zone
size with redshift, from which they inferred an order of magnitude
increase in the neutral fraction of the IGM over the observed range
$5.8\la z\la6.4$.  However, the detailed modeling of Bolton \&
Haehnelt~(2007a) showed that the evolution of the near-zone size is
not a simple function of the neutral hydrogen fraction. Their modeling
suggested that in the limit of a classical proximity zone the rapid
evolution of near-zone sizes could be explained by a relatively modest
evolution in the neutral fraction, with the rapid evolution being due
to the approach of the background transmission reaching 10 per cent.

The evolution of the ionization state of the IGM is thus difficult to
extract from the observed evolution of the near-zone size. In this
section we begin by discussing a calculation of quasar near-zone sizes
using our semi-analytic reionization model. We then describe the
observed relation between near-zone size and redshift before comparing
our model with the observed evolution of quasar near-zone size.

\subsection{Semi-analytic estimates for the evolution of near-zone sizes}

\begin{figure*}
\includegraphics[width=15.cm]{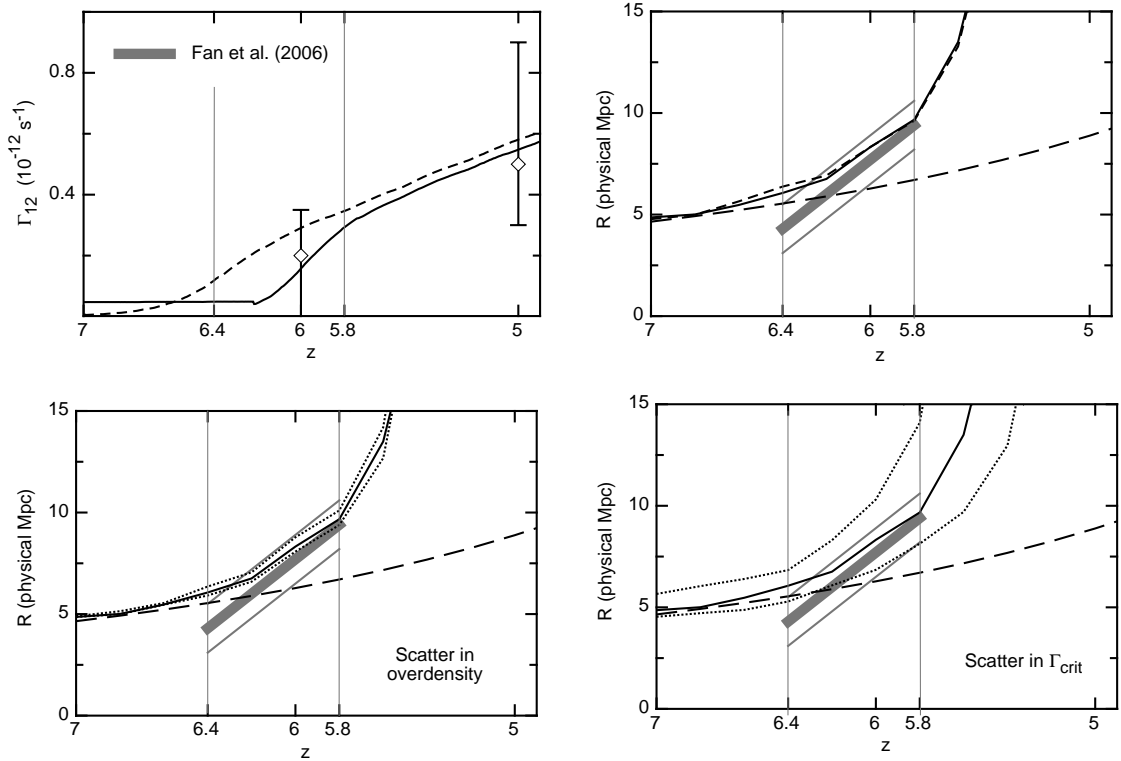}
\caption{\label{fig5} The predicted evolution of quasar near-zone
sizes as a function of redshift.  {\em Upper Left Hand Panel:} The
evolution of the ionization rate corresponding to regions at the mean
IGM density (solid line $\Delta_{\rm c}=20$; dashed line $\Delta_{\rm
c}=5$). {\em Upper Right Hand Panel:} The predicted evolution of the
near-zone size.  The solid line corresponds to the size expected
assuming the mean overdensity surrounding a quasar.  The long-dashed
line shows the near-zone evolution expected due to expansion of the
Universe only. Also shown for comparison (thick grey line) is the
observed near-zone size evolution, which may be parameterised by
$R_{\rm max}=7.7-8.1(z-6)$ physical Mpc (see section 4.2).   The thin
grey lines illustrate the level of observed scatter ($\pm1.2$Mpc).
{\em Lower Left Hand Panel:} As before, except now dotted curves
correspond to the near-zone size evolution computed at $-1\sigma$ and
$+1\sigma$ fluctuations around the mean density for $\Delta_{\rm
c}=20$.  {\em Lower Right Hand Panel:} As before, but now the dotted
lines correspond to the near-zone size evolution computed for
$z_{10}=5.0$ and $5.5$, each for $\Delta_{\rm c}=20$.  The overlap
redshifts (with {\fstar}=0.0037) of the mean IGM in these models are at
$z=6.2$ (with $\Delta_{\rm c}=20$) and  $z=7$  (with $\Delta_{\rm
c}=5$).  In all panels, the quasar host masses were assumed to be
$10^{13}M_\odot$, and quasar ionizing luminosities were
$\dot{N}=2\times10^{57}$s$^{-1}$.}
\end{figure*}

In Figure~\ref{fig4} we plot the ionization rate due to galaxies in a
biased, overdense region of the IGM surrounding a quasar.  The
ionization rates are displayed as a function of radius from the
quasar. The overlap of the mean IGM in the model used (which was shown
in Figures~\ref{fig1} and \ref{fig4} and has $\Delta_{\rm c}=20$) is
at $z=6.2$. The quasar host mass is $10^{13}M_\odot$, and its ionizing
luminosity is $\dot{N}=2\times10^{57}$s$^{-1}$. In the left-hand
panel, we show the ionization rate due to galaxies calculated at the
mean overdensity found within spheres of radius $R$ surrounding a high
redshift quasar.  The curves are plotted at a range of redshifts
between $z=5.6$ and $z=6.4$. The ionization rate due to the quasar
emission assuming the IGM to be optically thin is also shown (thick
grey line). We see that at redshifts near $z\sim5.8$, the value of the
ionization rate at distances corresponding to the observed quasar
near-zone sizes ($\sim10$Mpc) is comparable to the ionization rate due
to the quasar alone.  Moreover, at this epoch, the value of the galaxy
ionization rate increases rapidly with redshift due to the increasing
mean-free-path at the end of the overlap epoch. In the right hand
panel of Figure~\ref{fig4} we show the corresponding ionization rates
due to the sum of the galaxies plus the central quasar.  Note that 
we have linearly added the ionization rates and thus do
not account for the expected increase of the mean free-path due to 
the ionizing radiation of the quasar. We therefore still underestimate 
the ionization rate somewhat in the quasar near-zone in our model.  
The contribution to the ionization rate from galaxies flattens the radial
power-law profile of the ionization rate at large distances
surrounding the quasar. This flattening becomes more pronounced
towards lower redshifts.

We now turn to an analytical estimate of quasar near-zone sizes within
our reionization model. Adopting the near-zone size definition of Fan
et al.~(2006) discussed previously, we firstly need to estimate the
ionization rate which leads to a transmission level of 10 per cent.
Fan et al.~(2006) smoothed their spectra on a 20\AA~ scale to compute
the near-zone size; an effective \lya~ optical depth of $\tau_{\rm
eff} = -\ln(0.1) = 2.3$ is therefore an appropriate measure to
estimate this quantity.  The IGM exhibits an effective optical depth
of $2.3$ around $z_{10}\sim5.2$ (Songaila 2004, Fan et al.~2006), at
which point our semi-analytic reionization model predicts a value we
shall denote as $\Gamma(5.2)$.  Therefore, the limiting value of the
ionization rate which will correspond to a transmission level of 10
per cent at some redshift is approximately given by

\begin{equation}
\Gamma_{\rm lim} = \Gamma(5.2)\left(\frac{1+z}{6.2}\right)^{9/2},
\end{equation}
since $\Gamma \propto (1+z)^{9/2}$ for a fixed optical depth (e.g. Bolton
\& Haehnelt~2007a). We may therefore estimate the near-zone size as
the value of $R$ at which the sum of the galaxy and quasar ionizing
rates in Figure~\ref{fig4} drops below $\Gamma_{\rm lim}$. Note that
while we are confident in our model's ability to estimate the
background ionization rate, our analytic estimate of near-zone size is
very approximate due to the neglect of several important effects. These
effects include the attenuation of the quasars ionizing flux in the
near-zone as well as the effect of discrete absorbers on the observed
near-zone size (our analytic model computes only the mean properties
of a clumpy IGM). We will address these issues using a numerical model
of radiative transfer through a realistic simulation of the high
redshift IGM in \S~\ref{sec:sims}.

The semi-analytic predictions for the evolution of quasar near-zone
sizes as a function of redshift are shown in the upper right panel of
Figure~\ref{fig5}.  The redshift evolution for sizes calculated at the
mean overdensity surrounding a quasar is plotted as the solid line.
We show two cases, $\Delta_{\rm c}=20$ (solid line) and $\Delta_{\rm
c}=5$ (dashed line). The corresponding evolution of $\Gamma_{12}$ is
compared to the observational data in the upper left panel of
Figure~\ref{fig5}. Figure~\ref{fig5} shows that the near-zone size
should not evolve much prior to the redshift of overlap where the
quasar dominates the ionizing flux in the near-zone. However, when the
ionization rate due to galaxies approaches the level necessary to
maintain a mean transmission of 10 per cent, the near-zone size starts
to evolve rapidly until eventually the definition of the near-zone
size introduced by Fan et al. (2006) breaks down.  This is because the
flux level in most of the spectrum eventually exceeds 10 per cent, and
the edge of the near-zone becomes indistinguishable from the \lya~
forest.

\subsection{Observations of the near-zone size}

\begin{figure}
\includegraphics[width=7.cm]{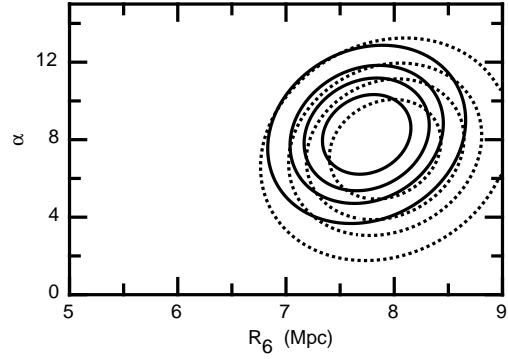}
\caption{\label{fig6} Constraints on the parameters $\alpha$ and
$R_6$ that describe the evolution of observed near-zone sizes with
redshift, assuming scalings of $L^{1/2}$ (dotted contours) and $L^{1/3}$ (solid contours)
respectively. In each case the extrema of the contours presented
represent the 68, 84, 91 and 97 per cent bounds  on single parameters.}
\end{figure}

We now discuss our predictions for the near-zone size evolution in
comparison with the observational data.  Fan et al.~(2006) found a
rapid evolution in the near-zone size with redshift. Their data show a
strong correlation about a mean relation, with a scatter that is in
excess of the uncertainty on individual parameters.  To remove the
dependence of near-zone size on quasar luminosity, $L$, Fan et
al.~(2006) rescaled their near-zone sizes to a common absolute
magnitude of $M_{1450}=-27$ by assuming the sizes are proportional to
$L^{1/3}$.  However, we caution that the uncertainty in the
interpretation of the nature of the near-zones (HII region {\it vs}
classical proximity zone) makes the scaling of the near-zone size with
the ionizing luminosity of the quasar rather uncertain (Bolton \&
Haehnelt~2007a).

In addition to the near-zone measurements of Fan et al.~(2006),
Willott et al.~(2007) have recently measured the near-zones for two
new high redshift quasars. In particular, they find sizes of 6.4Mpc at
$z=6.12$,  and 6.3Mpc at $z=6.43$. Willott et al.~(2007) use the
$L^{1/3}$ scaling suggested by Fan et al.~(2006), and obtain revised
values of 6.4Mpc at $z=6.12$ and  10.8Mpc at $z=6.43$, with the latter
point in particular apparently weakening the claim for evolution of
near-zone size with redshift (Willott et al.~2007). However Willott et
al.~(2007) used the definition of near-zone size suggested by Bolton
\& Haehnelt~(2007a), who defined the near-zone size to be the last
pixel at which the unsmoothed quasar spectrum drops below a
transmission level of 10 per cent.  This definition avoids smoothing
out important features in the spectrum, but breaks down once the IGM
as a whole becomes highly ionized and the edge of the near-zones
become ambiguous.  Here  we use the definition of
Fan et al. (2006) to facilitate a comparison with the 
the published sizes of Fan et al.
Inspection of the spectra published by Willott et al.~(2007) shows that the
near-zones are substantially smaller when the Fan et al.~(2006)
definition is employed. Therefore, our revised values for the
near-zone sizes of the Willott et al.~(2007) quasars are $\sim4.5$Mpc
at $z=6.12$ and $\sim3.5$Mpc at $z=6.43$. Rescaling these to account
for luminosity differences (using $L^{1/3}$) we find scaled values of
$\sim4.5$Mpc at $z=6.12$ and $\sim6$Mpc at $z=6.43$. Thus using a
consistent near-zone definition, the near-zone sizes for the new
quasars discovered by Willott et al.~(2007) are consistent with the
relation between near-zone size and redshift measured by Fan et
al.~(2006).

Following Fan et al.~(2006) we quantify the evolution of the near-zone
using a parameterised form for the evolution:
\begin{equation}
R = R_6 - \alpha (z-6),
\end{equation}
where $R_6$ is the value of the near-zone size at $z=6$, and $\alpha$
is the slope of the evolution in units of Mpc.  Assuming the
uncertainty ($\sigma_i$) for each of the $N_{\rm q}=18$
(including both the Fan et al.~(2006) and Willot et al.~(2007)
samples) observed near-zone sizes ($R_i$) at redshift $z_i$ to be
smaller than the intrinsic scatter ($\sigma_R$), we performed a
chi-squared fit minimising the variable
\begin{equation}
\chi^2(R_6,\alpha) = \Sigma_{i=0}^{i=N_{\rm
q}}\frac{[R_i-R(z_i)]^2}{\sigma_R^2}.
\end{equation}
The intrinsic scatter is estimated by adjusting $\sigma_{\rm R}$ to a
level that yields a minimum value of reduced-$\chi^2$ equal to unity.
 
The resulting constraints on $\alpha$ and $R_6$ are presented in
Figure~\ref{fig6} assuming scalings of $L^{1/3}$ (Fan et al.~2006) and
$L^{1/2}$, as is appropriate if the near-zone is the observational
signature of a classical proximity effect (Bolton \& Haehnelt~2007a).
The extrema of the contours shown  represent the 68, 84, 91 and 97
per cent bounds on single parameters. We find values of
$R_6=7.7\pm0.75$ and $\alpha=8.1\pm2$ with an intrinsic scatter of
$\sigma_R=1.2$ assuming a scaling of $L^{1/2}$ and values of
$R_6=7.9\pm1$ and $\alpha=7.5\pm2.5$ with an intrinsic scatter of
$\sigma_R=1.5$ assuming a scaling of $L^{1/2}$.  The resulting
constraint on $R_6$ is consistent with that quoted by Fan et
al.~(2006). However their quoted slope of $\alpha=9$ was somewhat
higher than our revised value, partly due to neglect of intrinsic
scatter, and partly due to the addition of the two additional quasars
from Willott et al.~(2007).

In the case of a uniformly ionized IGM, Bolton \& Haehnelt~(2007a)
have emphasised that the size of the quasar near-zone is independent
of both the neutral fraction of the surrounding IGM and the quasar's
lifetime, providing the background ionization rate at the near-zone
edge is small in comparison to the quasar ionization rate.  The
near-zone size defined in this way is plotted as the dashed line in
Figure~\ref{fig5} using the analytic model of Bolton \&
Haehnelt~(2007a) [their equation~(10) with $\Delta_{\rm
lim}=0.6$]. The figure shows that the evolution of near-zone size due
to density evolution alone would yield a value of $\alpha=2$, which is
excluded by the data at more than 95 per cent. This disagreement
implies that an additional effect is responsible for the observed
evolution. As suggested by Bolton \& Haehnelt (2007a), and as we shall
argue in more detail here using a model consistent with recent
observational constraints, this additional effect is most likely due
to the increasing amplitude of the ionizing background towards lower
redshifts which raises the overall transmission the IGM.

\subsection{Comparison between the semi-analytic model and observations}

For comparison with our semi-analytic model we show the observed
evolution of the quasar near-zones with redshift (Figure~\ref{fig5};
thick grey line). This evolution may be parameterised by $R_{\rm
max}=7.7-8.1(z-6)$ physical Mpc (Figure~\ref{fig6}).  The thin lines
illustrate the level of observed scatter. Both the predicted size and
evolution of the quasar near-zones are similar to observations for
this reionization model (which was not tuned beyond the requirement to
fit the ionization rate at $z\sim4-6$).

The near-zone sizes measured by Fan et al.~(2006) show significant
scatter, whose source we now discuss.  First, while quasars are
typically found in overdense regions, there are fluctuations in the
overdensity of the nearby IGM, and hence also fluctuations in the
stage of reionization occurring at a fixed redshift around different
quasars. In the lower left panel of Figure~\ref{fig5} we re-plot the
$\Delta_{\rm c}=20$ case for the average overdensity surrounding a
quasar, and also plot the curves corresponding to the $-1\sigma$ and
$+1\sigma$ density fluctuations around the mean (dotted lines).  Our
model predicts an amount of scatter in the near-zone sizes due to this
cosmic variance that is significantly smaller than the scatter in
observed near-zone sizes. The observed scatter is therefore due to the
differing properties of the density field along different quasar
lines-of-sight as has been shown previously in numerical simulations
(Bolton \& Haehnelt~2007a; Maselli et al. 2007; Lidz et al.~2007).
The variation among lines-of-sight, which must be described by a full
numerical treatment is manifested as scatter in the value of
ionization rate that allows a transmission of 10 per cent. We may
therefore estimate the scatter in the near-zone size using our
analytic model by computing the dependence of near-zone size on
redshift over the range of $z_{10}$ allowed by observation. This is
shown in the lower-right panel of Figure~\ref{fig5}, again for the
$\Delta_{\rm c}=20$ case, with the mean shown by the solid line and
the evolution for $z_{10}=5.0$ and $z_{10}=5.5$ shown by the dotted
lines. The scatter introduced is comparable to that of the
observations [note that we do not expect additional scatter from
quasar luminosity which has been scaled out in the relation of Fan et
al.~(2006), or from quasar lifetime which does not effect the size of
the near zone in a highly ionized IGM (Bolton \& Haehnelt~2007a)].
There may, however, be additional scatter due to variations in the
temperature of the surrounding IGM attributable to the temperature
dependence of the recombination coefficient.

Finally, it is important to note that the qualitative behaviour of the
near-zone size evolution shown in Figure~\ref{fig5} is independent of
the details of the reionization model. The two cases shown in
Figure~\ref{fig5} have overlap redshifts for the mean IGM at $z=6.2$
(with $\Delta_{\rm c}=20$) and $z=7$ (with $\Delta_{\rm c}=5$).  The
corresponding ionization rate as a function of redshift for these
models is shown in the upper left-hand panel of
Figure~\ref{fig5}. While the evolution of the near-zone sizes can be
successfully reproduced using an evolution for the ionization rate
predicted by our model at the tail-end of the overlap phase, the
near-zone sizes themselves cannot not be used as a probe of this
process.  This is because the rapid evolution in the near-zone sizes
is attributable to a relatively small change in the IGM effective
optical depth, and therefore is no more sensitive to the reionization
history than the Gunn-Peterson trough.  It is nevertheless gratifying
that we can  qualitatively reproduce the evolution of the
near-zone sizes with our analytical model, which is calibrated with
the evolution of the background ionization rate measured from the
effective optical depth in typical regions of the spectra.  
Note, however, that our analytic model is likely to overestimate the near-zone size
both because it does not account for attenuation of quasar flux due to
intervening neutral clumps in the IGM, and because the near-zone size
is computed for average absorption properties in a clumpy IGM. Full
numerical modeling with radiative transfer is required to reliably
interpret the spectra of individual high redshift quasars.  We
present such numerical simulations in the next section. 

\section{Modeling the observed near-zone size evolution with
radiative transfer simulations}
\label{sec:sims}

We now combine our semi-analytical model for the ionization rate in
the biased regions surrounding quasars with a radiative transfer
implementation and realistic density distributions drawn from a large
cosmological hydrodynamical simulation.  These simulations have been
discussed in detail in Bolton \& Haehnelt~(2007a), and their
description is not reproduced here.  However, we have made three
important changes. Firstly, we include an evolving, density dependent
ionizing background using our semi-analytic model, which has been
calibrated to observations at lower redshift.  This addition is
important because we have argued that it is the increase in the
intensity of the ionizing background which leads to the rapid increase
observed in the quasar near-zone sizes at $z<6$.  Secondly, we
construct the absorption spectra using an ionizing background computed
as a function of proper time along the trajectory of a photon emitted
by the quasar, rather than at the proper time of the quasar. This
effect is appropriate when considering spectra at the end of the
reionization era, when the ionizing background can evolve
significantly during the light travel time across a quasar near-zone.
Thirdly we have increased the temperature in the surrounding IGM
within 15 proper Mpc of the quasar to $\sim 40~000\rm~K$ to take
account the possibility that the hydrogen and helium in the near-zone
has been highly ionized  by a hard spectrum. We will discuss this
point in more detail below.

\begin{figure}
\includegraphics[width=8.5cm]{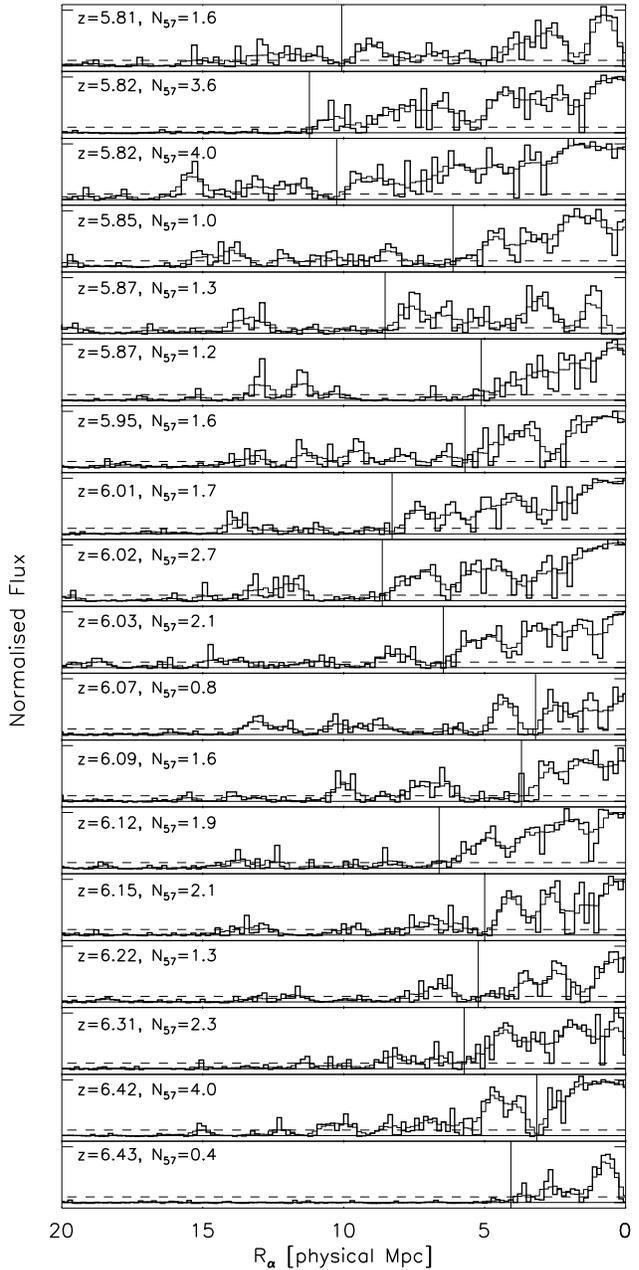}
\caption{\label{fig7} Synthetic Echelle Spectrograph
and Imager (ESI) spectra of high redshift quasar near-zones.  The
thick solid lines show the spectra at the computed resolution and the
thin solid lines show the spectra smoothed by a top-hat filter of
width 20\AA~ bins to allow comparison with observational data. The
vertical lines mark the measured near-zone size using the definition
of Fan et al.~(2006).}
\end{figure}

In Figure~\ref{fig7} we show the synthetic quasar absorption spectra
used in this study. The redshifts and luminosities  correspond to the
16 quasars analysed in the sample of Fan et al.~(2006), plus the two
additional quasars reported recently by Willott et al.~(2007).  The
emission rate of ionizing photons by the quasars, $\dot N$, is
displayed in each panel of Figure~\ref{fig7}, normalised by
$10^{57}\rm~s^{-1}$.   The thick solid lines show the synthetic
spectra after processing them to resemble data obtained with the Keck
telescope Echelle Spectrograph and Imager [see Bolton \&
Haehnelt~(2007a) for details], while the underlying thin solid lines
show these spectra after smoothing by a top-hat filter of width of
20\AA. The vertical lines mark the locations of the near-zone sizes as
defined by Fan et al.~(2006), where the smoothed spectrum first drops
below a transmission threshold of 10 per cent.  The spectra clearly
show the suppression of flux blueward of the \lya~ line, and the
gradual appearance of a Ly$\alpha$ forest towards lower redshifts.
The use of an evolving ionization rate calculated at the proper time
along the observed line-of-sight is necessary to consistently model
this progression.

\begin{figure}
\includegraphics[width=8.cm]{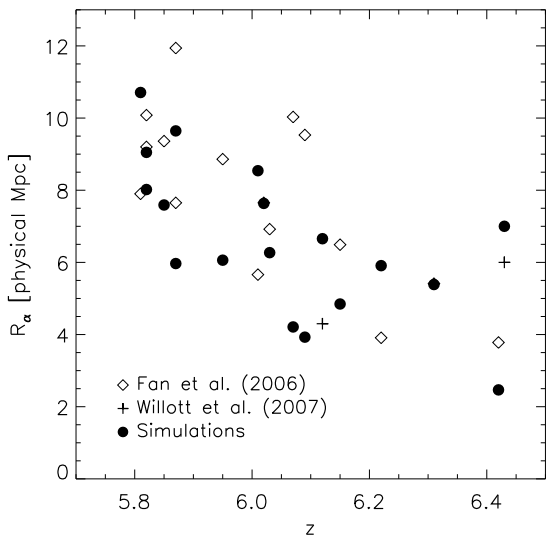}
\caption{\label{fig8}  The near-zone sizes measured from synthetic
spectra, plotted as a function of redshift (filled circles), along
with the observational data of Fan et al.~(2006) (open diamonds) and
Willott et al.~(2007) (crosses). The temperature in the surrounding
IGM within 15 proper Mpc of the quasar has been raised to $\sim
40~000\rm~K$ to take account the possibility that the helium in the
near-zone has been highly ionized by a hard spectrum. }
\end{figure}

\begin{figure}
\includegraphics[width=7.cm]{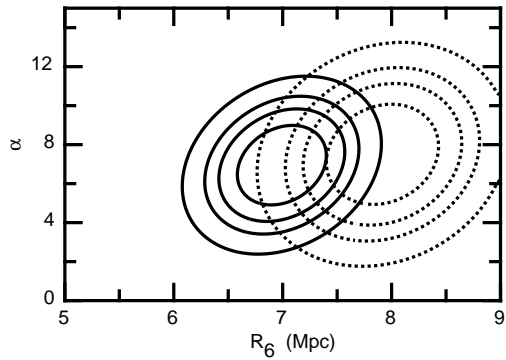}
\caption{\label{fig9}Constraints on the parameters $\alpha$
and $R_6$ that describe the evolution of model near-zone sizes with
redshift. The model had $\dot{N}=2\times10^{57}$s$^{-1}$ and a
large IGM temperature within 15Mpc of the quasar.  In each case the
extrema of the contours presented represent the 68, 84, 91 and 97 per cent
bounds  on single parameters.  The observational constraints assuming
a scaling of $L^{1/2}$ are reproduced (dotted contours).}
\end{figure}

The sizes of the near-zones derived from the 18 synthetic spectra are
plotted as a function of redshift in Figure~\ref{fig8} (filled
circles) along with the observational data from Fan et al.~(2006)
(open diamonds) and Willott et al.~(2007) (crosses).  Following Fan et
al.~(2006), the sizes have been rescaled to a common AB magnitude of
$M_{1450}=-27$ by assuming the near-zone size is proportional to $\dot
N^{1/3}$. There is very good agreement between the observed and
simulated sample of spectra. We have performed the $\chi^2$ minimisation on the
parameterised evolution for the sample of 18 simulated near-zones,
with the resulting constraints plotted in Figure~\ref{fig9}.

We find parameters describing the mean relation of $R_6=7.0\pm0.4$ and
$\alpha=6.9\pm2$. The intrinsic scatter is at a level of
$\pm1.2$Mpc. These near-zones are slightly smaller than those
observed. The modeled near-zones show an evolution that is consistent
with the observed rate of evolution. However as we found for the
observed near-zones, the models are not consistent with evolution due
to IGM density alone ($\alpha=2$).  The scatter of model near-zone
sizes is consistent with the scatter among the observed near-zone
sizes.  Note, however, that any scatter due to expected variations
in the quasar emission histories and the subsequent impact on the
IGM has been neglected in the simulations.

We now return to the discussion regarding the temperature of the IGM
we adopted when constructing the synthetic near-zone spectra.  When
computing $\dot N$, each quasar is assumed to have a broken power law
spectral energy distribution (SED) similar to that reported by Telfer et
al.~(2002),

\begin{eqnarray}
\nonumber \epsilon_{\nu} &\propto&  \nu^{-0.5}\hspace{3mm}
\mbox{for}\hspace{3mm}1050<\lambda<1450\,\AA,\\ &\propto&\nu^{-1.5}
\hspace{3mm}\mbox{for}\hspace{3mm}\lambda<1050\,\AA
\end{eqnarray}

\noindent Thus, $M_{1450}=-27$ corresponds to
$\dot{N}=1.9\times10^{57}$s$^{-1}$, and the 18 quasars in the Fan et
al.~(2006) and Willott et al.~(2007) samples span a range of
$0.4<\dot{N}/10^{57}\rm s^{-1}<4.0$.  In Bolton \& Haehnelt (2007a) it
was noted that, even with a suitable increase in the ionizing
background towards lower redshift, these quasar luminosities appeared
to be too low to reproduce the absolute sizes of the observed
near-zones with the simulations when using the definition advocated by
Fan et al. (2006), particularly at $z<6$ where the near-zone sizes are
increased by the additional transmission from the emerging \lya~forest.

Consequently, in order to reproduce the observational result of Fan et
al. (2006), each quasar in the simulations of Bolton \& Haehnelt
(2007a) also had a value of $\dot N$ which was increased by a factor of
$2.5$ relative to the value corresponding to $M_{1450}=-27$ for our
adopted SED.  Since a factor of $2.5$ increase in the quasar
luminosity is rather large, it was argued that this increase may be
instead attributable to the increased mean free path for ionizing
photons in the near-zone, which leads to an enhanced contribution from
the ionizing background due to galaxies.  In this work we have
consistently modeled the ionizing background due to galaxies by
including the effect of reionization bias in the near-zone in detail.
Nevertheless, we find the enhancement of the ionization rate in the
near-zone due to this effect alone is still not enough to resolve this
difference in the simulations.  Note that this is not in contradiction
with our analytical model predictions for the near-zone sizes.  As
stressed previously, the analytical estimates are likely to
overestimate the near-zone sizes.  We have therefore re-examined the
assumptions which go into our numerical modeling.

In the case of a classical proximity zone the neutral hydrogen
fraction and thus the sizes of the near-zones do not only depend on
the ionization rate but are also rather sensitive to the temperature
of the IGM in the near-zone.  This is due to the temperature
dependence of the recombination rate for ionized hydrogen; in
ionization equilibrium the neutral hydrogen density $n_{\rm HI}\propto
T^{-0.7}/\Gamma_{12}$.  Therefore, a factor of $2.5$ increase in the
ionization rate is equivalent to around a factor of four increase in
the ambient IGM temperature.  In Bolton and Haehnelt (2007a) the
numerical simulations had a median IGM temperature of $\sim
15~000\rm~K$ before the quasar turned on, with the gas assumed to be
in ionization equilibrium with the ionizing background.  However,
since the cooling timescale for the low density IGM is typically of
order a Hubble time at $z=6$, changes in the gas temperature made
during a putative earlier accretion phase by the quasar would still
persist.  For example, if the quasar reionized the hydrogen and helium
around it during an earlier phase when it had a very hard spectral
index and the IGM was largely neutral, radiative transfer effects can
increase the temperature in the IGM to $\sim 30~000- 40~000\rm~K$
(Abel \& Haehnelt~1999; Bolton, Meiksin \& White~2004; Tittley \&
Meiksin~2006).  The IGM will then have only cooled marginally during
the interval between active quasar phases.  Although our radiative
transfer scheme does correctly include these kind of effects, the
heating boost is not present in our current simulations.  This is
because we assume the IGM is already highly ionized by our model
ionizing background before the quasar turns on, and we do not model
the previous emission history of the quasar, which would require
substantially more detailed simulations.

Therefore, to mimic this effect, we have raised the temperature in our
simulations within 15 Mpc of the quasar by a factor of $2.5$,
corresponding to a median value of $\sim 40~000\rm ~K$. Note that this
temperature is smaller than the value of $\sim 60~000\rm~K$ one would
expect based according to the scaling relation discussed above, which
is higher than can be easily achieved by photoionization. Nevertheless
the near-zone sizes are in good agreement with that of the observed
sample.  The deviation from the expected scaling is due to the
reionization bias which we had not taken into account in  Bolton \&
Haehnelt (2007a). Our model for the increased ionizing background in
the biased regions surrounding quasars helps to increase the near-zone
sizes and drive their rapid evolution.  As discussed above, our
simulations provide a good description of the observed near-zone sizes
with the observed luminosities and reasonable assumptions for the
quasar SED, and may thus provide a hint that the IGM is substantially
hotter than average around the highest redshift quasars.

There remains a small discrepancy between observed and modeled
near-zone sizes (Figure~\ref{fig9}). However, as already mentioned we
have not modeled the effect of the quasar radiation on the mean free
path in the near-zone. There is also the possibility that the
discrepancy  is partially due to a continuum on the observed spectra
which has been placed too low, or perhaps a quasar SED which is
different to the one we have assumed.  Considering these uncertainties
the agreement appears excellent.

In summary, our semi-analytic modeling of the history for the ionizing
background, combined with numerical simulation of the Ly$\alpha$
transmission in quasar near-zones, is able to quantitatively describe
the evolution of quasar near-zone sizes at $6.4<z<5.8$, and the
evolution of the ionizing background at $z\sim4-6$.  We find that the
rapid evolution observed in the sizes of the high redshift quasar
near-zones is due to the increasing level of ionizing background near
$z\sim6$, with absolute sizes which suggest an increased IGM
temperature in the vicinity of the quasars.

\section{SUMMARY AND CONCLUSIONS}
\label{conclusions}

Absorption spectra of high redshift quasars show an increasingly thick
Ly$\alpha$ forest, suggesting an increase in the fraction of neutral
gas in the IGM approaching $z\sim6$, culminating in the complete lack
of detected flux in the Gunn-Peterson trough blueward of the \lya~
line. However the meaning of the complete Gunn-Peterson troughs
remains controversial, having been interpreted both as the completion
of the reionization epoch, and as merely a gradual thickening of the
Ly$\alpha$ forest. The inconclusive nature of the observations has its
root in the very large cross-section for absorption of \lya~ photons,
which allow a direct probe of the neutral hydrogen content of the IGM
only down to volume averaged neutral fractions of around $10^{-4}$.

The fact that quasars reside in massive halos places them in overdense
and strongly biased regions of the IGM. This presents an additional
challenge for the interpretation of regions in quasar absorption
spectra close to the quasars intrinsic redshift, since one is using
results from these biased regions to infer the properties of the mean
IGM. In this work we have presented a model for the evolution of the
emissivity of ionizing photons and the ionization state of the IGM
that takes such a reionization bias in the environment of bright,
high-redshift quasars into account.  In our CDM based model the
ionizing photons for reionization are produced by star formation in
dark matter halos spanning a wide mass range. At redshifts both prior
to, and post overlap, our model allows us to calculate  values for the
volume and mass-averaged  neutral fractions, the ionizing photon
mean-free-path and the evolution of the ionizing background. With
appropriate choices for the parameters controlling the star formation
efficiency, we are able to reproduce the ionizing emissivity at
$4<z<6$ and the ionizing photon mean-free-path at $z\sim4$ inferred
from the observed opacity  distribution in \lya forest absorption
spectra. In our model the comoving ionizing emissivity is roughly
constant with redshift and the volume filling factor of ionizing
regions increases slowly from 10 to 90 per cent between $z\ga10$
and $z\sim 6-7$, where reionization completes following the overlap of
cosmological HII regions.  Within 5 physical Mpc of a high redshift
quasar, we find the evolution of the ionization state of the IGM
precedes that of the mean universe by around 0.3 redshift
units. During the period shortly after overlap the mean-free-path is
seen to increase rapidly in the model.  The offset in the redshift of
overlap between overdense regions and the mean IGM therefore results
in an ionizing background near high redshift quasars (excluding the
flux of ionizing radiation from the quasar itself) that exceeds the
value  in the mean IGM by a factor of $\sim2-3$ for a short period of
time.

High redshift quasar spectra also show evidence for a highly ionized
region immediately surrounding the quasar.  Using our semi-analytic
prescription for biased reionization, we model the evolution of these
quasar near-zones.  We find near-zone sizes which exhibit a redshift
dependence which is qualitatively consistent with the current
observational data.  The model indicates that the rapid evolution in
the near-zone sizes in the redshift range $5.8<z<6.4$ is due to the
increasing background ionization rate originating from an almost
constant galactic emissivity.  Within our model, this rise in the
background ionization rate is therefore due to the rapid rise in the
mean free path for ionizing photons, which is expected near the
tail-end of the reionization epoch as \HII~ regions overlap.

In order to model the opacity distribution in spectra of bright
high-redshift quasars in more detail we have combined our model for
the evolution of the metagalactic ionization rate in the biased
environment surrounding quasars with detailed line-of-sight radiative
transfer simulations in the inhomogeneous high-redshift IGM.  In this
way we were able to reproduce the sizes of the observed near-zones and
their rapid evolution in the redshift range $5.8<z<6.4$ very well. Our
numerical modeling also suggests that the rapid evolution is not due
directly to large changes in the neutral hydrogen fraction of the IGM.
Instead, as suggested by Bolton \& Haehnelt (2007a), the near-zone
sizes probe the evolution of the background ionization rate, which
results in modest changes to an already very small neutral fraction.
We furthermore find that the absolute sizes of the near-zones are
rather sensitive to the IGM temperature in the near-zones, due to the
temperature dependence of the recombination rate.  The observed
near-zone sizes are reproduced given the observed quasar ionizing
luminosities if the temperature of the surrounding IGM is high ($\sim
40~000\rm~K$).  Heating to such high temperatures at distances of up
to $10-15$ proper Mpc may be plausible if the hydrogen and helium in
the near-zone has been highly ionized by a hard spectrum prior to the
completion of reionization in the IGM as a whole.  This requires the
continuous or intermittent emission of a hard spectrum for an extended
period of time.  If the IGM temperature is not as hot as
$40~000\rm~K$,  the differences found between modeled and observed
near-zone sizes could be explained as at least partially being due to
a continuum on the observed spectra which has been placed too low,  or
perhaps a quasar SED which is different to the one we have assumed.

There are still several issues to address regarding the sizes and
redshift evolution of high redshift quasar near-zones.  Larger samples
will  clarify the significance of the evolution observed in the
current data and hopefully constrain the evolution at even higher
redshift. A more detailed modeling of the quasar accretion history and
the ensuing flux of ionizing photons may be required to
self-consistently model near-zone sizes.  The detection of strongly
broadened absorption lines within quasar near-zones  would lend
support to a scenario in which IGM temperatures within quasar
near-zones are increased by the reionization of hydrogen and helium by
a hard spectrum.   However, in this work we have presented a simple
reionization model with a roughly constant ionizing emissivity, in
which reionization ends following a rapid increase in the ionizing
photon mean-free-path as \HII~ regions overlap at $z\sim 6-7$.  This
scenario successfully reproduces the gross features present in
high redshift quasar absorption spectra,  both within and outside of
the quasar near-zones.

\section*{Acknowledgments}
This work was supported in part by the Australian Research Council.
JSBW acknowledges the hospitality of the Institute of Astronomy at
Cambridge University where part of this work was undertaken.

\end{document}